\documentclass[11pt]{article}
\usepackage{latexsym,amsmath}
\usepackage{epsfig}

\def\bg#1{\mbox{\boldmath$#1$}}

\newcommand{\beq}{\begin{eqnarray}}
\newcommand{\eeq}{\end{eqnarray}}
\newcommand{\be}{\begin{eqnarray*}}
\newcommand{\ee}{\end{eqnarray*}}

\newcommand{\ra}{\rightarrow}

\newcommand{\nn}{\nonumber}

\newcommand{\bo}{{\bg\omega}}
\newcommand{\bO}{{\bg\Omega}}
\newcommand{\bd}{{\bf d}}
\newcommand{\bR}{{\bf R}}
\newcommand{\inv}[1]{\frac{1}{#1}}

\begin{document}

\centerline{\large\bf{Dynamics of the scalar field in 5-dimensional Kaluza-Klein theory}}
\vskip 5mm
\centerline{Ingunn Kathrine Wehus and Finn Ravndal}
\vskip 5mm
\centerline{\it Institute of Physics, University of Oslo, N-0316 Oslo, Norway.}

\begin{abstract}

\footnotesize{
Using the language of differential forms, 
the Kaluza-Klein theory in 4+1 dimensions is derived.
This theory unifies electromagnetic and gravitational 
interactions in four dimensions when the extra space dimension is
compactified. Without any ad-hoc assumptions
about the five-dimensional metric, the theory also contains a scalar
field coupled to the other 
fields. By a conformal transformation the theory is transformed from the Jordan frame to the Einstein frame
where the physical content is more manifest. Including a cosmological constant in the five-dimensional
formulation, it is seen to result in an exponential potential for the scalar field in four dimensions. 
A similar potential is also found from the Casimir energy in the compact dimension. The resulting 
scalar field dynamics mimics realistic models recently proposed for cosmological quintessence.}

\end{abstract}

\section{Introduction}
A unified formulation of Einstein's theory of gravitation and Maxwell's theory of electromagnetism in
four-dimensional spacetime was first proposed by Kaluza\cite{Kaluza} by assuming a pure gravitational theory 
in a five-dimensional spacetime. The metric components were to be independent of the fifth 
coordinate\cite{Nordstrom}. This so-called 'cylinder condition' was a few years later explained by Klein
by invoking arguments from the newly established quantum mechanics when the extra dimension was compactified
on a circle $S^1$ with a microscopic radius\cite{Klein}. In the following years it was
studied by a large number of authors and also extended to higher dimensions in order to incorporate non-Abelian
gauge theories\cite{KK}. In the last couple of years it has seen a rebirth as the mathematical implementation
of the exciting possibility that our Universe can have extra dimensions which in principle can be of macroscopic
size\cite{A}\cite{ADD}\cite{AADD}\cite{HLZ}\cite{GRW}.

The field content of the original Kaluza-Klein theory is given by the five-dimensional metric $\bar{g}_{MN}$
where the indices $M,N = 0,1,2,3,4$. Denoting the indices of our four-dimensional spacetime by Greek indices, 
its metric is given by the components $\bar{g}_{\mu\nu}$ while the electromagnetic vector
potential is given by the components $\bar{g}_{\mu 4} = \bar{g}_{4\mu}$. The remaining spatial component 
$\bar{g}_{44}$ was for no good reason set equal to a constant by Kaluza\cite{Kaluza} and kept that way
by the subsequent authors. This assumption 
was apparently first abandoned by Jordan\cite{Jordan} and then later
by Thiry\cite{Thiry}
who showed that $\bar{g}_{44}$ corresponds to a scalar field in our spacetime. At that time there was no need or place for
such a field, but today we realize that it is a generic feature of almost any extension of Einstein's theory
of gravity. From the experimental point of view it can be related to the physics behind the dark energy\cite{PR}
in the Universe in the form of a cosmological constant\cite{SW} or variable quintessence\cite{DW}.

The main motivation behind the present contribution is to investigate some of the physical properties
and manifestations of such a scalar field. In order to make the presentation more accessible, we present the
reduction of the five-dimensional theory down to our four-dimensional spacetime in more detail than is generally
available in the literature. We will take advantage of the simplifications which follow from using differential 
forms as was first done by Thiry\cite{Thiry}. 

After this reduction, we find that the scalar field couples directly
to the gravitational field. To get this part on the standard form, we perform a conformal transformation so that
the gravitational part is described by the usual Einstein-Hilbert action. In this frame the scalar field
couples only to the Maxwell field. It is shown that this form of the
full action can can be obtained
by performing a conformal transformation on the five-dimensional metric. The classical equations of motion for 
the different fields are then derived and discussed.

If there is a cosmological constant in the five-dimensional spacetime, we show that it gives rise to an
exponential potential for the scalar field. Such potentials are phenomenologically very interesting in
cosmology where they can drive the observed acceleration of the Universe\cite{DW}. A potential of the same
form also follows from the Casimir energy in the compactified, extra dimension. Including
both of these effects, we have an effective potential which is the sum of two exponential terms. Adjusting
the parameters in this potential, one can then obtain a realistic cosmology of an accelerating 
Universe\cite{Cope}\cite{Maju}.

\section{Reduction to four dimensions}

In the five-dimensional spacetime\footnote{The flat metric in this spacetime has the diagonal components 
$\eta_{MN} = (-1,1,1,1,1)$.} we have the line element $d\bar{s}^2 = \bar{g}_{MN}dx^M dx^N$ where the
coordinates are $x^M = (x^\mu, x^4\equiv y)$. It is convenient to define $\bar{g}_{44} = h$ and
$\bar{g}_{4\mu} = hA_\mu$ where the fields $h(x)$ and $A_\mu(x)$ at this stage depend on all five
coordinates. Our four-dimensional spacetime is orthogonal to the basis vector $\vec{e}_4$ in the extra
direction. It is therefore spanned by the four basis vectors 
\beq
       \vec{e}_{\mu\perp} = \vec{e}_\mu -\vec{e}_{\mu\parallel}
     = \vec{e}_\mu - \frac{\vec{e}_\mu\cdot\vec{e}_4}{\vec{e}_4\cdot\vec{e}_4}\vec{e}_4
     = \vec{e}_\mu - \frac{\bar{g}_{\mu 4}}{\bar{g}_{44}}\vec{e}_4                               \label{basisv}
\eeq
and is endowed with the metric
\beq
     g_{\mu\nu} =  \vec{e}_{\mu\perp}\cdot\vec{e}_{\nu\perp}
                =  \bar{g}_{\mu\nu} - \frac{\bar{g}_{4\mu}\bar{g}_{4\nu}}{\bar{g}_{44}}
               =   \bar{g}_{\mu\nu} - h A_{\mu}A_{\nu}                                         \label{4metric}
\eeq
In this way we then have the following splitting of the five-dimensional metric
\beq
       \bar{g}_{MN}= \left[\begin{array}{c}
                              \begin{tabular}{c | c}
                    $g_{\mu\nu}+ h A_{\mu}A_{\nu}$ & $h A_{\mu}$ \\
                    \hline
                    $h A_{\nu}$ & $h$ \\
                \end{tabular}
         \end{array}\right]                                                                    \label{g_MN}
\eeq
and correspondingly for the contravariant components
\beq 
    \bar{g}^{MN}= \left[\begin{array}{c}
                              \begin{tabular}{c | c}
                    $g^{\mu\nu}$ & $ -A^\mu$ \\
                    \hline
                    $ - A^\nu$ & $h^{-1} + A_\mu A^\mu$ \\
                \end{tabular}
         \end{array}\right]
\eeq
which satisfy $\bar{g}_{MN}\bar{g}^{NP}=\delta_M^{\;\;P}$. The line element in this spacetime thus takes
the form
\beq
      d\bar{s}^2 &=& \bar{g}_{\mu\nu}dx^{\mu} dx^{\nu} + 2\bar{g}_{\mu 4}dx^{\mu} dx^4 
           + \bar{g}_{44}dx^4 dx^4 \nn \\
           &=& g_{\mu\nu}dx^{\mu} dx^{\nu} + h\left(dy + A_{\mu}dx^{\mu}\right)^2           \label{ds2}
\eeq 
The appearance of the gauge potential 1-form in the last term is characteristic for Kaluza-Klein theories
also with additional extra dimensions\cite{KK}. At this stage we impose the cylinder condition
$\bar{g}_{MN,4} = 0$ which means that the fields $g_{\mu\nu}(x), A_\mu(x)$ and $h(x)$ are independent of
the fifth coordinate $y$.

\subsection{Vierbeins and connection forms}

In the four-dimensional spacetime we introduce the vierbeins $V^{\hat\mu}_{\;\;\nu}$ and their inverse 
$V^\mu_{\;\;\hat\nu}$ satisfying  $V^{\hat\mu}_{\;\;\lambda}V^\lambda_{\;\;\hat\nu} = 
\delta^{\hat\mu}_{\;\hat\nu}$ and $V^\mu_{\;\;\hat\nu}V^{\hat\nu}_{\;\;\lambda} = \delta^\mu_{\;\lambda}$.
The metric is thus $g_{\mu\nu} = V^{\hat\alpha}_{\;\;\mu}V^{\hat\beta}_{\;\;\nu}\eta_{\hat\alpha\hat\beta}$ where
$\eta_{\hat\alpha\hat\beta}$ is the four-dimensional Minkowski metric in this spacetime. Using these vierbeins we 
find the orthonormal basis 1-forms as 
\beq
\bo^{\hat\mu} =
V^{\hat\mu}_{\;\;\nu}\,\bd x^\nu                              \label{omij}
\eeq
In the fifth direction
we see from the line element (\ref{ds2}) that the basis 1-form is 
\beq
      \bo^{\hat 4} = \sqrt{h}(\bd y + A_\mu\,\bd x^\mu)                                       \label{om4}
\eeq
Now using standard methods\cite{MTW}, we can calculate the connection 1-forms and the curvature 2-forms
needed to find the Riemann and Ricci tensors which enter the Einstein-Hilbert action. As an example,
we find 
\be
     \bd\bo^{\hat 4} = {1\over 2}h^{-1/2}h,_\mu\bd x^\mu\wedge (\bd y + A_\mu\,\bd x^\mu) 
                    + h^{1/2}A_{\mu,\rho}\,\bd x^\rho\wedge \bd x^\mu
\ee
since $\bd^2y = 0$. 
The result can be written in the form 
$\bd\bo^{\hat 4}= -{\bar\bO}^{\hat4}_{\;\;\hat\mu}\wedge\bo^{\hat\mu}$
where ${\bar\bO}^{\hat4}_{\;\;\hat\mu}$ is the corresponding five-dimensional connection 1-form. 
By using the antisymmetry of $\bd x^\rho\wedge \bd x^\mu$ to introduce 
the field strength $F_{\mu\nu} = A_{\nu,\mu} - A_{\mu,\nu}$, we  find 
\beq
       {\bar\bO}^{\hat4}_{\;\;\hat\mu} = {1\over 2}\sqrt{h} F_{\hat\mu\hat\nu}\,\bo^{\hat\nu}
                                          + {1\over 2h}h,_{\hat\mu} \bo^{\hat 4}
\eeq
where $h,_{\hat\mu} = h,_\nu V^\nu_{\;\;\hat\mu}$. Using the antisymmetry in the two orthonormal indices, we then 
also have the value for ${\bar\bO}^{\hat\mu}_{\;\;\hat 4}$ which we will need in the following.

The exterior derivative of the remaining basis forms is similarly given in five dimensions as
\be
     \bd\bo^{\hat\mu} = -{\bar\bO}^{\hat\mu}_{\;\;\hat\nu}\wedge\bo^{\hat\nu}   
                        -{\bar\bO}^{\hat\mu}_{\;\;\hat 4}\wedge\bo^{\hat 4} 
\ee
while taken in the four-dimensional spacetime it would just be $\bd\bo^{\hat\mu} 
= -\bO^{\hat\mu}_{\;\;\hat\nu}\wedge\bo^{\hat\nu}$ where $\bO^{\hat\mu}_{\;\;\hat\nu}$ now is the the
connection 1-form in four dimensions. We thus have the relationship 
\beq
        {\bar\bO}^{\hat\mu}_{\;\;\hat\nu} = \bO^{\hat\mu}_{\;\;\hat\nu} 
                                          + {1\over 2}\sqrt{h}F_{\hat\nu}^{\;\;\hat\mu}\bo^{\hat 4}
\eeq
between the connections in these two spacetimes.

\subsection{Curvature forms and tensors}

The curvature 2-forms are defined by the structure equation
\beq
      {\bR}^{\hat\mu}_{\;\;\hat\nu} = \bd{\bO}^{\hat\mu}_{\;\;\hat\nu} 
   + {\bO}^{\hat\mu}_{\;\;\hat\lambda}\wedge{\bO}^{\hat\lambda}_{\;\;\hat\nu} 
   \equiv {1\over 2}R^{\hat\mu}_{\;\;{\hat\nu}{\hat\rho}{\hat\sigma}}\bo^{\hat\rho}\wedge \bo^{\hat\sigma} 
\eeq 
in the four-dimensional spacetime and correspondingly in five dimensions. Their components
${R}^{\hat\mu}_{\;\;{\hat\nu}{\hat\rho}{\hat\sigma}}$ form the Riemann curvature tensor in the
orthonormal frame we are working in. In the same frame the corresponding expansion of the connection forms is
\beq
      \bO^{\hat\mu}_{\;\;\hat\nu} = \Omega^{\hat\mu}_{\;\;\hat\nu\sigma}\bd x^\sigma
\eeq
where $\Omega^{\hat\mu}_{\;\;\hat\nu\sigma}$ are the connection coefficients. These will enter the calculation
together with partial derivatives of the field strengths $F^{\hat\mu}_{\;\;\hat\nu}$ to give the covariant
derivative
\beq
     F^{\hat\mu}_{\;\;\hat\nu ;\sigma} =  F^{\hat\mu}_{\;\;\hat\nu ,\sigma} 
                                       + \Omega^{\hat\mu}_{\;\;\hat\rho\sigma}F^{\hat\rho}_{\;\;\hat\nu} 
                                       - \Omega^{\hat\rho}_{\;\;\hat\nu\sigma}F^{\hat\mu}_{\;\;\hat\rho}
\eeq  
and similarly for the other components.

We calculate first the curvature form ${\bar\bR}^{\hat\mu}_{\;\;\hat 4}$ from
\beq
      {\bar\bR}^{\hat\mu}_{\;\;\hat 4} = \bd{\bar\bO}^{\hat\mu}_{\;\;\hat 4} 
    + {\bar\bO}^{\hat\mu}_{\;\;\hat\nu}\wedge{\bar\bO}^{\hat\nu}_{\;\;\hat 4} 
\eeq  
Along the direction $\bo^{\hat\rho}\wedge \bo^{\hat\sigma}$ we then find the curvature tensor components
\beq
     {\bar R}^{\hat\mu}_{\;\;{\hat 4}{\hat\rho}{\hat\sigma}} 
                          = {1\over 2}\sqrt{h} F_{\hat\sigma\hat\rho}^{\;\;\;\;;\hat\mu}
                          + {1\over 4\sqrt{h}}\left(2h^{;\hat\mu}F_{\hat\sigma\hat\rho}
                          + h_{;\hat\rho}F_{\hat\sigma}^{\;\;\hat\mu}
                          - h_{;\hat\sigma}F_{\hat\rho}^{\;\;\hat\mu}\right)
\eeq
and similarly
\beq
    {\bar R}^{\hat\mu}_{\;\;{\hat 4}{\hat\rho}{\hat 4}} = {1\over 4}F^{\hat\mu\hat\nu}F_{\hat\rho\hat\nu}
                          - {1\over 2h}h^{;\hat\mu}_{\;\;\; ;\hat\rho}
                          + {1\over 4h^2}h^{;\hat\mu}h_{;\hat\rho}
\eeq
in the direction of $\bo^{\hat\rho}\wedge \bo^{\hat 4}$. These are all the components of the Riemann tensor 
involving the fifth index, since the Riemann tensor is antisymmetric in the first two and in the last two indices. We have
here introduced the notation  $F_{\hat\sigma\hat\rho ;\hat\mu} = F_{\hat\sigma\hat\rho ;\nu}
V^\nu_{\;\;\hat\mu}$ for the covariant derivative of the field tensor. 
We have also simplified the result using the Bianchi identity $F_{\hat\sigma\hat\rho ;\hat\mu}
+F_{\hat\rho\hat\mu;\hat\sigma} + F_{\hat\mu\hat\sigma ;\hat\rho} = 0$.

The remaining components of the curvature tensor follow now from the 2-form
\beq
      {\bar\bR}^{\hat\mu}_{\;\;\hat\nu} = \bd{\bar\bO}^{\hat\mu}_{\;\;\hat\nu} 
    + {\bar\bO}^{\hat\mu}_{\;\;\hat\lambda}\wedge{\bar\bO}^{\hat\lambda}_{\;\;\hat\nu} 
    + {\bar\bO}^{\hat\mu}_{\;\;\hat 4}\wedge{\bar\bO}^{\hat 4}_{\;\;\hat\nu} 
\eeq  
which can be evaluated along the same lines. It gives
\beq
      {\bar R}^{\hat\mu}_{\;\;{\hat\nu}{\hat\rho}{\hat\sigma}} = R^{\hat\mu}_{\;\;{\hat\nu}{\hat\rho}{\hat\sigma}}
      + {1\over 4}h\left(2F_{\hat\nu}^{\;\;\hat\mu}F_{\hat\rho\hat\sigma}
      - F_{\hat\rho}^{\;\;\hat\mu}F_{\hat\sigma\hat\nu} - F_{\hat\sigma}^{\;\;\hat\mu}F_{\hat\nu\hat\rho}\right)
\eeq
which has the correct antisymmetry in the first and last two indices. We also notice that the symmetry
${\bar R}_{\hat\mu\hat\nu\hat\rho\hat\sigma} = {\bar
R}_{\hat\rho\hat\sigma\hat\mu\hat\nu}$ is satisfied, as well as ${\bar R}^{\hat\mu}_{\;\;[{\hat\nu}{\hat\rho}{\hat\sigma}]}=0$.

The Ricci curvature tensor is defined in four dimensions to be $R_{\hat\nu\hat\sigma} = 
R^{\hat\mu}_{\;\;\hat\nu\hat\mu\hat\sigma}$ and similarly in five
dimensions. It is symmetric in its two indices. We find its components
to be
\beq
     {\bar R}_{\hat4\hat4} &=& {1\over 4}hF^{\hat\mu\hat\nu}F_{\hat\mu\hat\nu}
      -  {1\over 2h}h^{;\hat\mu}_{\;\;\; ;\hat\mu}  +  {1\over 4h^2}h^{;\hat\mu}h_{;\hat\mu}      \label{R44} \\
       {\bar R}_{\hat\nu \hat4} &=& {1\over
     2}\sqrt{h}F_{\hat\nu\hat\mu}^{\;\;\;\; ;\hat\mu}
                                  + {3\over 4\sqrt{h}}F_{\hat\nu\hat\mu}h^{;\hat\mu}   \label{R4i} \\
       {\bar R}_{\hat\nu\hat\sigma} &=& R_{\hat\nu\hat\sigma} 
                                     -  {1\over 2}h F^{\hat\mu}_{\;\;\hat\nu}F_{\hat\mu\hat\sigma} 
                                     - {1\over 2h}h_{;\hat{\nu}\hat{\sigma}}
                                     + {1\over 4h^2}h_{;\hat\nu}h_{;\hat\sigma}      \label{Rij} 
\eeq
For the scalar curvature
${\bar R} = {\bar R}^{\hat{\mu}}_{\;\;\hat{\mu}}$ expressed in a coordinate basis we then have
\beq
     {\bar R} = R - {1\over 4}hF^{\mu\nu}F_{\mu\nu} - {1\over h}\nabla^2 h 
              +  {1\over 2h^2}(\nabla_{\mu}h)^2                                                  \label{Ricci}
\eeq
in agreement with Thiry\cite{Thiry}.
Here we have introduced the $\nabla$-operator for the covariant derivative and $\nabla^2 =\nabla^{\mu}\nabla_{\mu}$ is 
the four-dimensional d'Alembertian operator. 
Needless to say, it is the appearance of the Maxwell Lagrangian in
this higher-dimensional curvature first derived by Kaluza and Klein, that we still don't understand
the full significance of.

\subsection{Einstein-Hilbert action and equations of motion}

These geometrical considerations become physical when we postulate that gravitation in the five-dimensional
space is governed by the corresponding Einstein-Hilbert action
\beq
      S = {1\over 2}{\bar M}^3\int\!d^5x\sqrt{-{\bar g}} {\bar R}                          \label{EH5}
\eeq
where ${\bar M}$ is the Planck mass in this space and ${\bar R}$ is
the Ricci curvature scalar (\ref{Ricci}).
In principle there can also be an additional term here corresponding to a cosmological constant. Its 
implications will be considered in section \ref{koskonst}. From the $4+1$ split of the metric in (\ref{g_MN}) we see
that its determinant is simply ${\bar g} = h g$ where $g$ is the determinant of the four-dimensional metric
$g_{\mu\nu}$. Using this in the action (\ref{EH5}) and then integrating out the fifth coordinate, we find
the action
\beq
     S = {1\over 2}M^2\int\!d^4x\sqrt{-g}\sqrt{h}\left[R - {1\over 4}hF^{\mu\nu}F_{\mu\nu} - {1\over h}\nabla^2 h 
       + {1\over 2h^2}(\nabla_{\mu}h)^2\right]                                   \label{EH4}
\eeq
when the extra dimension is a microscopic circle of radius $a$ so that
$M^2 = {2\pi a{\bar M}^3}$ becomes the
ordinary, four-dimensional Planck constant. The two last terms form a total derivative,
\beq
     \sqrt{h}\left[{1\over h}\nabla^2 h - {1\over 2h^2}(\nabla_\mu h)^2\right]  
  = 2\nabla^2\sqrt{h}                                                                       \label{surf}
\eeq
Assuming that $h$ disappears far away these terms can be
neglected from the action. We are therefore left with the final result
\beq
     S = {1\over 2}M^2\int\!d^4x\sqrt{-g}\sqrt{h}\left[R - {1\over 4}hF^{\mu\nu}F_{\mu\nu}\right]   \label{KK}  
\eeq
which is the Kaluza-Klein action. 

In the general case where the scalar field $h(x)$ varies with position, the effective gravitational constant 
given by the coefficient of the Ricci scalar in (\ref{KK}), is no longer a constant, but varies with time and 
position in the four-dimensional spacetime. Electromagnetic interactions described by the Maxwell part, will 
similarly have a variable coupling strength. For this reason the theory seems to be in disagreement with
present-day observations although there have been recent indications that the fine-structure constant may 
vary over cosmological time scales\cite{alpha}. Kaluza-Klein theory thus belongs to a wider class of fundamental theories 
characterized by the extension of Einstein's tensor theory of gravity to include also the effect of scalar
interactions. Such scalar-tensor theories of gravitation were constructed by Jordan\cite{Jordan_1} and 
later shown by Brans and Dicke\cite{BD} to be compatible with gravitational experiments and cosmological tests.

The classical equations of motion for the three fields can be derived from the action (\ref{KK}). But it is
simpler to use the five-dimensional action (\ref{EH5}) which gives rise to the equation of motion ${\bar R}_{MN} = 0$.

We have already the components of the Ricci tensor in orthonormal basis
(\ref{R44}-\ref{Rij}). We now transform these to coordinate basis
using the vierbeins $V^{\hat\mu}_{\;\;\nu}$, along with the rest of the
f\"{u}nfbein components $V^{\hat\mu}_{\;\;4}=0$,
$V^{\hat4}_{\;\;\mu}=\sqrt{h} A_{\mu}$ and $V^{\hat4}_{\;\;4}=\sqrt h $, 
which we read out from (\ref{omij}) and (\ref{om4}). 
We then find 
\begin{align}
        {\bar R}_{44} =& 
{1\over 4}h^2 F^{\mu\nu}F_{\mu\nu} -{1\over 2}\nabla^2 h  +  {1\over
4h}(\nabla_{\mu}h)^2   \label{R44k} 
\\
        {\bar R}_{\mu 4} =& 
{1\over 2}{h}\nabla^{\nu}F_{\mu\nu}+{3\over4}F_{\mu\nu}\nabla^{\nu}h
+A_{\mu} \left[
{1\over 4}h^2 F^{\mu\nu}F_{\mu\nu} -{1\over 2}\nabla^2 h  +  {1\over
4h}(\nabla_{\mu}h)^2 
\right]         \label{R4ik} 
\\
\begin{split}
       {\bar R}_{\mu\nu} =& R_{\mu\nu} - {1\over 2}h F^{\sigma}_{\;\;\mu}F_{\sigma\nu}
                                     - {1\over 2h}\nabla_{\mu}\nabla_{\nu}h
                                     +{1\over4h^2}(\nabla_{\mu}h)(\nabla_{\nu}h) 
\\ \phantom{=}&
+A_{\mu}A_{\nu}\left[{1\over 4}h^2 F^{\mu\nu}F_{\mu\nu}      
-  {1\over 2}\nabla^2 h  +  {1\over 4h}(\nabla_{\mu}h)^2 \right] 
\\  \phantom{=}& 
+A_{\nu}\left[{1\over 2}{h}\nabla^{\sigma}F_{\mu\sigma}
+{3\over4}F_{\mu\sigma}\nabla^{\sigma}h\right]
+A_{\mu}\left[{1\over 2}{h}\nabla^{\sigma}F_{\nu\sigma}+
{3\over4}F_{\nu\sigma}\nabla^{\sigma}h\right]
\end{split}     \label{Rijk} 
\end{align}
By equating these components of the Ricci scalar to zero
we find the corresponding four-dimensional equations of motion. 
With the help of the identity (\ref{surf}), we find
from (\ref{R44k}) for the scalar field 
\beq
      \nabla^2\sqrt{h} = {1\over 4} h^{3/2}F^{\mu\nu}F_{\mu\nu}                    \label{h_eq}
\eeq
while (\ref{R4ik}) combined with (\ref{h_eq}) gives the equation of motion for the Maxwell field,
\beq\label{A_eq}
     \nabla^{\mu}F_{\mu\nu} = -{3\over 2h}F_{\mu\nu}\nabla^{\mu}h
\eeq
Finally, for the gravitational field we find from (\ref{Rijk}) when
using (\ref{h_eq}) and (\ref{A_eq})
\beq\label{g_eq}
      R_{\mu\nu} = {1\over 2} h F^{\sigma}_{\;\;\mu}F_{\sigma\nu} + {1\over\sqrt{h}}\nabla_\mu\nabla_\nu\sqrt{h}
\eeq
These equations can also be found in Wesson \cite{Wesson}. 
Introducing the Einstein curvature tensor $E_{\mu\nu} = R_{\mu\nu} 
- {1\over 2}Rg_{\mu\nu}$, we see that the last equation can be written
as
\beq\label{E_eq}
    E_{\mu\nu} = {1\over 2}h \Big(F^{\sigma}_{\;\;\mu}F_{\sigma\nu} 
               - {1\over 4}g_{\mu\nu}F^{\rho\sigma}F_{\rho\sigma}\Big) 
+{1\over\sqrt{h}}
\Big(\nabla_\mu\nabla_\nu\sqrt{h}-g_{\mu\nu}\nabla^2\sqrt{h}\Big)
\eeq
when we express the Ricci scalar $R$ in terms of the Maxwell and scalar fields using (\ref{Ricci}) with
${\bar R} = 0$. 
In the first term we recognize the energy-momentum tensor of the electromagnetic field, while the
last term must be the corresponding entity for the scalar field in this representation.

From the equation of motion (\ref{h_eq}) for the scalar field we see that it can take a constant value, which
can be chosen to be $h = 1$, provided the accompanying gauge field satisfies the special condition 
$F^{\mu\nu}F_{\mu\nu} = 0$. The magnitudes of the electric and magnetic fields must therefore be the same 
everywhere. This rather strong and unnatural condition was imposed for many years in investigations of 
the Kaluza-Klein theory\cite{KK} since there didn't seem to be a real physical need for a scalar field on the 
same footing as the electromagnetic and gravitational fields. Today, however, the situation is different. In fact, scalar 
fields are at the core of the Higgs mechanism in particle physics, cosmological inflation in the early universe
and dark energy in the late universe. In the following we will therefore keep the scalar field non-constant and
study some of its physical implications.

\section{Conformal transformations to the Einstein frame}

A basic assumption in Einstein's general theory of relativity is that all observers are equipped with standard
measuring rods and clocks. The properties of these rods and clocks are coded into the components of the metric tensor $g_{\mu\nu}$
and has the consequence that the gravitational action is just given by the volume integral of the Ricci 
curvature scalar. This is obviously the case for the underlying, five-dimensional theory described by (\ref{EH5}).
But in the resulting, four-dimensional theory (\ref{KK}) we see that this term is multiplied by the scalar field
$\sqrt{h}$ which is generally not constant. This corresponds to using non-standard measuring rods and clocks. We
can now adjust these by changing the metric at every point by a Weyl transformation
\beq
     g_{\mu\nu} \ra \Omega^2  g_{\mu\nu}                                                      \label{Weyl_g}
\eeq
and choosing the scale factor $\Omega(x)$ appropriately. In a $D$-dimensional spacetime this results in the
corresponding change
\begin{equation}\label{Ricciweyl}
\begin{split}
R_{\mu\nu}\ra R_{\mu\nu}-&\Omega^{-1}\nabla^2\Omega g_{\mu\nu}
-(D-2)\Omega^{-1}\nabla_{\mu}\nabla_{\nu}\Omega \\
-&(D-3)\Omega^{-2}(\nabla_{\rho}\Omega)^2 g_{\mu\nu}
+2(D-2)\Omega^{-2}\nabla_{\mu}\Omega\nabla_{\nu}\Omega
\end{split}
\end{equation}
of the Ricci tensor. This gives a change in the Ricci scalar
\beq
       R \ra \Omega^{-2}R - 2(D-1)\Omega^{-3}\nabla^2\Omega 
                      - (D-1)(D-4)\Omega^{-4}(\nabla_\mu\Omega)^2                             \label{Weyl_R}
\eeq
as showed in \cite{HE}.

\subsection{Weyl transformations}

When we are in $D=4$ dimensions $\sqrt{-g}\ra\Omega^4\sqrt{-g}$ while
the Ricci scalar changes
according to (\ref{Weyl_R}). The first term in
(\ref{KK}) thus changes as 
\beq
     \int\!d^4x\sqrt{-g}\sqrt{h} R 
     \ra  \int\!d^4x\sqrt{-g}\,\Omega^4\sqrt{h}\,\Omega^{-2}R + \ldots                         \label{trans_4}
\eeq
For the coefficient of $R$ to take the canonical value we must therefore choose $\Omega = h^{-1/4}$. Including
the Maxwell term in (\ref{KK}) and the second term of the Weyl transformation (\ref{Weyl_R}) we thus find the
transformed Kaluza-Klein action
\beq
     S = {1\over 2}M^2\int\!d^4x\sqrt{-g}\left[R - {1\over 4}h^{3\over 2}F^{\mu\nu}F_{\mu\nu}
       + {3\over 2}{1\over h}\nabla^2 h - {15\over 8}{1\over h^2}(\nabla_{\mu}h)^2 \right]     \label{Weyl_KK}  
\eeq
It can be simplified by combining the last two terms by a partial integration which results in
\beq
     S = {1\over 2}M^2\int\!d^4x\sqrt{-g}\left[R - {1\over 4}h^{3\over 2}F^{\mu\nu}F_{\mu\nu}
       - {3\over 8}{1\over h^2}(\nabla_{\mu}h)^2 \right]                                       \label{Weyl_KK1}  
\eeq
After this Weyl transformation we are now in the Einstein frame. By construction the gravitational part in the 
first term of the action has now the canonical form. The last term describes a massless scalar field which is 
coupled to the electromagnetic field in the second term. This is the
physical content of the Einstein frame theory. For similar transformations between
the Jordan frame and the Einstein frame for dilatonic brane-worlds see\cite{NOOT}. 

We could have achieved the same result by performing the Weyl transformation (\ref{Weyl_g}) directly on the 
five-dimensional metric appearing in (\ref{EH5}). 
Since we now have $\sqrt{-\bar{g}}\ra \Omega^5\sqrt{-\bar{g}}$, we
find that the transformation needed is $\Omega=h^{-1/6}$. 
Again using (\ref{Weyl_R}) where now also the last term 
contributes and the gradients act in five dimensions, we find
\beq
     \int\!d^5x\sqrt{-\bar g}{\bar R} 
     \ra  \int\!d^4x\sqrt{-g}\,\left[{\bar R} 
    + {4\over 3}{1\over {h'}}\nabla^2 { h'} 
                - {1\over { h'}^2}(\nabla_{\mu}{h'})^2 \right]           \label{trans_5}
\eeq
For later convenience we have here denoted the scalar field $h'$ instead of $h$. By using
(\ref{Ricci}) for the five-dimensional scalar curvature, we find the
transformed action integral
\beq
     S = {1\over 2}M^2\int\!d^4x\sqrt{-g}\left[R - {1\over 4}{h'} F^{\mu\nu}F_{\mu\nu}
       + {1\over 3}{1\over {h'}}\nabla^2 {h'} 
       - {1\over 2}{1\over {h'}^2}(\nabla_{\mu}{h'})^2 \right]                        \label{Weyl_KK2}  
\eeq
Again we can use a partial integration to combine the two last terms,
as in (\ref{Weyl_KK1}). The final result for the Kaluza-Klein
action after a five-dimensional Weyl transformation is then
\beq
     S = {1\over 2}M^2\int\!d^4x\sqrt{-g}\left[R - {1\over 4}{h'} F^{\mu\nu}F_{\mu\nu}
       - {1\over 6}{1\over {h'}^2}(\nabla_{\mu}{h'})^2 \right]                        \label{Weyl_KK3}  
\eeq
We see that there is full agreement between the four-dimensional Weyl
transformation $g_{\mu\nu}\ra h^{-1/2}g_{\mu\nu} $ and the five-dimensional
counterpart $\bar{g}_{MN}\ra h^{-1/3}\,\bar{g}_{MN}$. 
If we put $h' = h^{3/2}$ in (\ref{Weyl_KK3}) this equation is transformed
into (\ref{Weyl_KK1}). 

This is also easily understood directly from the five-dimensional metrical structure (\ref{g_MN}). After the
four-dimensional Weyl transformation the five-dimensional metric becomes
\be
       \left[\begin{array}{c}
                              \begin{tabular}{c | c}
                    $h^{-{1\over 2}}g_{\mu\nu}+ h A_{\mu}A_{\nu}$ & $h A_{\mu}$ \\
                    \hline
                    $h A_{\nu}$ & $h$ \\
                \end{tabular}
         \end{array}\right]     =  h^{-{1\over 2}}\left[\begin{array}{c}
                              \begin{tabular}{c | c}
                    $g_{\mu\nu}+ h^{3\over 2} A_{\mu}A_{\nu}$ & $h^{3\over 2} A_{\mu}$ \\
                    \hline
                    $h^{3\over 2} A_{\nu}$ & $h^{3\over 2}$ \\
                \end{tabular}
         \end{array}\right]
\ee
Introducing here ${h'} = h^{3/2}$ we then have the 
Weyl transformation of the five-dimensional metric
used above.

\subsection{Canonical fields}

Although the gravitational part of the action now has the standard form, the kinetic energies of scalar and 
electromagnetic fields do not have their canonical forms. However, this is simple to achieve. In the action
(\ref{Weyl_KK3}) we introduce
\beq
             { h'} = e^{\sqrt{6}\phi/M}                                                \label{can_phi}
\eeq
where now $\phi(x)$ is a scalar field with canonical normalization. Similarly, we redefine the electromagnetic 
field by 
\beq\label{can_A}
A_\mu \ra \sqrt{2}A_\mu/M
\eeq
 and the Kaluza-Klein action takes its final form
\beq
     S = \int\!d^4x\sqrt{-g}\left[{1\over 2}M^2 R - {1\over 4}e^{\sqrt{6}\phi/M} F^{\mu\nu}F_{\mu\nu}
       - {1\over 2}(\nabla_\mu\phi)^2\right]                                                 \label{Weyl_KK4}  
\eeq
When the scalar field $\phi$ has values much less than the Planck mass
$M$, both the electromagnetic field and the scalar field are
seen to be free.

\subsection{Equations of motion in Einstein frame}

Now that we have performed the Weyl transformation $\bar{g}_{MN}\ra
h^{-\inv{3}} \bar{g}_{MN}$ 
the equations of motion are
no longer the same as the equations (\ref{h_eq}), (\ref{A_eq})
and (\ref{E_eq}) 
in the Jordan frame. We can find the
new equations of motion from the transformed four-dimensional
action (\ref{Weyl_KK3}). Varying this action with respect to $h$
gives
\begin{equation}
\nabla^2 h=\frac{3}{4} h^2 F_{\mu\nu}F^{\mu\nu}+\frac{1}{h} (\nabla_{\mu}h)^2 
\label{bev1} 
\end{equation}
Similarly the equation for $A_{\mu}$ is found to be
\begin{equation}
\nabla^{\nu}F_{\mu\nu}=-\frac{1}{h} F_{\mu\nu}\nabla^{\nu}h
\label{bev2}  
\end{equation}
while varying with respect to $g_{\mu\nu}$ gives us
\begin{equation}
E_{\mu\nu}={1\over2}\left[ hT_{\mu\nu}^{\text{el.mag.}}+{1\over{3h^2}} T_{\mu\nu}^{\text{scalar}}\right]
\label{bev3}
\end{equation}
Here $T_{\mu\nu}^{\text{el.mag.}}
=F^{\alpha}_{\;\;\mu}F_{\alpha\nu}-\frac{1}{4}F^{\rho\sigma}F_{\rho\sigma}g_{\mu\nu}$ is the ordinary electromagnetic
energy-momentum tensor,
while $T_{\mu\nu}^{\text{scalar}}=\nabla_{\mu}h\nabla_{\nu}h-\frac{1}{2}
(\nabla_{\sigma}h)^2g_{\mu\nu}$  is the energy-momentum
tensor for a free scalar field. 
By performing the canonical transformations (\ref{can_phi}) and (\ref{can_A})
we then obtain the equations of motion in canonical normalization:
\beq
\nabla^2 \phi &=&\sqrt{\frac{3}{8M^2}} e^{\sqrt{6}\phi/M}F_{\mu\nu}F^{\mu\nu} 
\label{bev1c} \\
\nabla^{\nu}F_{\mu\nu}&=&-\frac{\sqrt{6}}{M} F_{\mu\nu}\nabla^{\nu}\phi
\label{bev2c}  \\
E_{\mu\nu}&=&{1\over{M^2}}\left[e^{\sqrt{6}\phi/M}
T_{\mu\nu}^{\text{el.mag.}}+ T_{\mu\nu}^{\text{scalar}}\right]
\label{bev3c}
\eeq

We can also get the equations (\ref{bev1}-\ref{bev3}) by
varying the five-dimensional action after the Weyl
transformation. This gives ${\bar{R}}'_{MN}=0$, where
${\bar{R}}'_{MN}$ is the transformed Ricci tensor which we find from
(\ref{Ricciweyl}).
In our case $\Omega^2=h^{-\inv{3}}$ and we  get the following
expression for the transformed Ricci tensor: 
\begin{equation}\label{Riccitrans}
{\bar R}'_{MN}
=\bar{R}_{MN}-\frac{1}{4h^2}(\bar{\nabla}_{\!P}h)^2 \bar g_{MN}
-\frac{5}{12h^2}\bar{\nabla}_{\!M}h\bar{\nabla}_{\!N}h
+\frac{1}{2h}\bar{\nabla}_{\!M}\bar{\nabla}_{\!N} h+\frac{1}{6h}\bar{\nabla}^2 h g_{MN}
\end{equation}
Here we have barred the covariant derivative to remind the reader that
it is the covariant derivative with respect to the five-dimensional metric. 
The relationship between the d'Alembertian operator in four and five dimensions is
$\bar{\nabla}^2 h = \nabla^2 h + \inv{2h}(\nabla_{\mu}h)^2$. 
By computing a couple of Christoffel symbols we find
\beq
\bar{\nabla}_{\!4}\bar{\nabla}_{\!4}h&=&\inv{2}(\nabla_{\mu}h)^2 \\
\bar{\nabla}_{\!4}\bar{\nabla}_{\!\nu}h&=&\inv{2}(\nabla_{\mu}h)^2 A_{\nu}-\inv{2}h
\nabla_{\mu}h F_{\nu}^{\;\mu} \\
\bar{\nabla}_{\!\mu}\bar{\nabla}_{\!\nu}h&=&\nabla_{\!\mu}\nabla_{\!\nu}h
+\inv{2}(\nabla_{}\sigma h)^2 A_{\mu}A_{\nu}-\frac{1}{2}h \nabla^{\rho}h
\left(A_{\mu}F_{\nu\rho}+A_{\nu}F_{\mu\rho}\right) 
\eeq
When we now equate the various components of ${\bar R}'_{MN}$ to
zero and use the expressions (\ref{R44k}-\ref{Rijk})
for the untransformed Ricci tensor, we end up with equations
(\ref{bev1}), (\ref{bev2}) and (\ref{bev3}). To get the last equation
we must also use the fact that the transformed Ricci scalar
${\bar{R}}'=0$. 


\section{Potential energy for the scalar field}

So far the scalar field is massless and will therefore modify the gravitational interactions over cosmological
distances. This is surely unwanted and is easily avoided by slightly enlarging the theory. By including a
cosmological constant in five dimensions we will see that the scalar field develops a potential and thus also
a non-zero mass. Alternatively, we will see that the Casimir vacuum energy induced by the presence of the
compact, fifth dimension also generates a similar potential.

\subsection{Cosmological constant in five dimensions}\label{koskonst}

With a cosmological constant $\bar\Lambda$ in the original, five-dimensional theory, the fundamental
action (\ref{EH5}) is replaced by
\beq
      S = {1\over 2}{\bar M}^3\int\!d^5x\sqrt{-{\bar g}}\,\big( {\bar R} - 2\bar\Lambda \big)     \label{LEH5}
\eeq
Going through the same compactification as before, followed by the Weyl transformation in five dimensions,
it immediately follows that
\beq
     S = {1\over 2}M^2\int\!d^4x\sqrt{-g}\left[R - {1\over 4}{h'} F^{\mu\nu}F_{\mu\nu}
       - {1\over 6}{1\over {h'}^2}(\nabla_{\mu}{h'})^2 
       - 2\bar\Lambda{h'}^{-{1\over 3}}\right]                                                \label{KKL}
\eeq
Introducing here the canonically normalized fields, it takes the more informative form
\beq
     S = \int\!d^4x\sqrt{-g}\left[{1\over 2}M^2 R - {1\over 4}e^{\sqrt{6}\phi/M} F^{\mu\nu}F_{\mu\nu}
       - {1\over 2}(\nabla_\mu\phi)^2 - M^2{\bar\Lambda}e^{-\sqrt{2\over 3}\phi/M}\right]      \label{KKL4}  
\eeq
The five-dimensional cosmological constant is thus seen to correspond to an exponential potential in four
dimensions. Its absolute sign is directly set by the sign of $\bar\Lambda$.

Such an exponential potential for a scalar field addition to Einstein's tensor theory was first considered by
Wetterich\cite{Wett}. It has been much studied since then in connection with models for 
quintessence\cite{DW}. Its cosmological evolution is completely characterized by the coefficient 
of $\phi$ in the exponent.

\subsection{Casimir energy from the compact dimension}

The cosmological constant is a contribution to the vacuum energy from physics on short scales, i.e. scales 
shorter than the size 
\beq
     L = \int_0^{2\pi a}\! dy \sqrt{h} = 2\pi a e^{\sqrt{2/3}\phi/M}                \label{L_comp}
\eeq
of the compact dimension. But including quantum effects, the Casimir energy will contribute to the vacuum 
energy at the scale $L$ due to the confinement of the massless field quanta in a space with a compact, fifth 
dimension. The corresponding momentum is therefore quantized with the values $p = (2\pi/L)n$ where 
$n = 0, \pm 1 ,\pm 2, \dots, \pm\infty$. In (\ref{L_comp}) 
we have chosen to calculate $L$ after the four-dimensional Weyl
transformation, which means that the correspondence between $\phi(x)$
and $h(x)$ is given by
$h= (e^{\sqrt{6}\phi/M})^{2/3}= e^{2\sqrt{2/3}\phi/M}   $.
The calculation of the Casimir energy is done in lowest order perturbation theory where the field quanta represent small oscillation
around the ground state $\bar{g}_{MN} = \eta_{MN}$ which follows from the 
classical equations of motion.

Considering first the contribution from one such massless mode, it gives rise to the Casimir energy
\beq
     E_0 = {V\over 2}\int\!{d^3k\over (2\pi)^3}\sum_{n= -\infty}^\infty\sqrt{k^2 + (2\pi n/L)^2}
\eeq
where $V$ is a finite 3-volume. Using dimensional regularization, we do the momentum integral in $d$ dimensions
where we have the general formula
\beq
        \int\!{d^dk\over (2\pi)^3}{1\over (k^2 + m^2)^N} 
       = (4\pi)^{-d/2}{\Gamma(N-d/2)\over\Gamma(N)}(m^2)^{d/2 - N}
\eeq
In our case $N= - 1/2$ and we find
\be
     E_0 = {V\over 2}\sum_{n= -\infty}^\infty (4\pi)^{-d/2}{\Gamma(-{d+1\over 2})\over\Gamma(-{1\over 2})}
           \Big({2\pi n\over L}\Big)^{d+1}
\ee
The sum over the compact quantum number $n$ is still divergent. It is made finite with zeta-function 
regularization which gives
\beq
       \sum_{n= -\infty}^\infty n^{d+1} = 2\zeta(-d-1)
\eeq
Taking now $d \ra 3$, we have
\beq
    E_0 = - {V\over L^4}\pi^2\Gamma(-2)\zeta(-4)                                     \label{E_01}
\eeq
Although $\Gamma(-2)$ is infinite and $\zeta(-4)$ is zero, their product is finite as follows from the
reflection formula
\beq
      \Gamma\Big({z\over 2}\Big)\pi^{-z/2}\zeta(z) 
     =\Gamma\Big({1- z\over 2}\Big)\pi^{-(1-z)/2}\zeta(1- z)
\eeq
for zeta-functions. 
We put $z= - 4$ in this formula and use it to rewrite $E_0$. The
corresponding vacuum energy density 
${\cal E}_0 = E_0/VL$ in five dimensions is then
\beq
     {\cal E}_0 = - {3\,\zeta(5)\over 4\pi^2L^5}                                        \label{E_02} 
\eeq
This is in agreement with the result of Applequist and Chodos\cite{Casimir} obtained by a more indirect approach. 
It also follows from the calculations of Ambj\o rn and Wolfram\cite{AW} who studied Casimir energies in 
different geometries. More recently the corresponding Casimir energy has been calculated by Albrecht, 
Burgess, Ravndal and Skordis\cite{ABRS} in a similar Kaluza-Klein theory in six dimensions of which two are 
compact in order to derive a corresponding scalar potential. Elizalde
et. al. have calculated the corresponding Casimir energy for
an anti-de Sitter background\cite{ENOO}.

In five dimensions the graviton has five physical degrees of freedom. The total 
Casimir energy is thus the above calculated result (\ref{E_02}) times
five. 
With the length $L$ expressed by the scalar 
field as in (\ref{L_comp}), we
put this energy into the five-dimensional action integral and obtain the potential
\beq
      V(\phi) = 5{\cal E}_0 \left(2\pi a\right)\sqrt{h}\left(h^{-{1\over 4}}\right)^4
              = - {15\,\zeta(5)\over (2\pi)^6a^4}e^{-6\sqrt{2\over 3}\phi/M}
\eeq
in the Einstein frame. 
This potential term enters the Kaluza-Klein action 
instead of the last term in (\ref{KKL4}). It has the same exponential form as the contribution from the 
cosmological constant, but the exponent is six times as large.

In general the potential will get contributions both from the small-scale cosmological constant and from the
Casimir energy. It will thus have the form
\beq
       V(\phi) = Ae^{-\alpha\phi}  +  Be^{-\beta\phi}
\eeq
where the exponents are fixed but the coefficients are unknown. We only know that at least one of the
prefactors $A$ or $B$ must be negative in this particular theory. 

Including also higher order quantum corrections to the Casimir energy, the simple exponential result will be 
modified. In the six-dimensional theory of Albrecht, Burgess, Ravndal and Skordis\cite{ABRS} the radiative 
corrections add up to a polynomial in the field $\phi$. This modification can be important when such potentials 
are used in models for cosmological quintessence\cite{AS}.

\section{Conclusions}

The original, five-dimensional theory of Kaluza and Klein is the simplest example of the more elaborate
theories used today to describe physics with extra dimensions. In addition to unifying the electromagnetic
and gravitational fields, it also contains a scalar field which codes the size of the single extra dimension
here. Similar theories with additional compact dimensions will contain a corresponding scalar field which is 
usually called the radion\cite{ADD}. This represents an extension of Einstein's tensor theory of gravity and
can have important, cosmological consequences. In this connection the
radion field appears as quintessence which can give
rise to acceleration of the Universe at late times. This depends on
the field dynamics which is
governed by its effective potential which appears in the Einstein frame. 

Here it is pointed out that this effective potential will 
in the lowest order approximation used here
be a sum of two exponential terms. One is resulting from the small-scale cosmological 
constant in the higher-dimensional spacetime while the other is induced as a Casimir energy due to 
one or more compact dimensions. It has been shown by others that potentials of such a form allow for a 
consistent description of the evolution of the Universe since radiation domination until today when the dark 
energy dominates and gives acceleration\cite{Cope}\cite{Maju}. It would be even more satisfactory if also 
the inflationary mechanism could be explained by similar physics from extra dimensions.


\begin{thebibliography}{99}

\bibitem{Kaluza} T. Kaluza, {\it Sitzungsber. Preuss. Akad. Wiss. Phys. Mat. Klasse,} 966 (1921).

\bibitem{Nordstrom} G. Nordstr\"om, {\it Phys. Zeit.} {\bf 15}, 504 (1914), had earlier used the same idea
        to formulate an in four dimensions unified theory of
        electromagnetism and his scalar theory of gravitation,  
        based on Maxwell's theory in a five-dimensional spacetime.

\bibitem{Klein} O. Klein, {\it Zeit. f. Phys.} {\bf 37}, 895 (1926), {\it Nature} {\bf 118}, 516 (1926).

\bibitem{KK} T. Appelquist, A. Chodos and P.G.O. Freund, {\it Modern Kaluza-Klein Theories}, Addison-Wesley
             Publishing Company, Menlo Park, California (1987).

\bibitem{A} I. Antoniadis, {\it Phys. Lett.} {\bf B246}, 377 (1990); 

\bibitem{ADD} N. Arkani-Hamed, S. Dimopoulos and G. Dvali, {\it Phys. Lett.} {\bf B429}, 263 (1998);
             {\it Phys. Rev.} {\bf D59}, 086004 (1999).

\bibitem{AADD} I. Antoniadis, N. Arkani-Hamed, S. Dimopoulos and
G. Dvali, {\it Phys. Lett.} {\bf B436}, 257 (1998); 

\bibitem{HLZ} T. Han, J. D. Lykken and R. Zhang, {\it Phys. Rev.} {\bf D59},  105006 (1999).

\bibitem{GRW} G.F. Giudice, R. Rattazzi and J.D. Wells, {\it Nucl. Phys.} {\bf B544}, 3 (1999).

\bibitem{Jordan} P. Jordan, {\it Ann. d. Physik} {\bf 1}, 219 (1947).

\bibitem{Thiry} Y. Thiry, {\it Comptes Rendus} {\bf 226}, 216 (1948).

\bibitem{PR} P.J.E. Peebles and B. Ratra, {\it astro-ph/0207347}. 

\bibitem{SW} S. Weinberg, {\it astro-ph/0005265}.

\bibitem{DW} For a recent summary, see M. Doran and C. Wetterich, {\it astro-ph/0205267}.

\bibitem{Cope} T. Barreiro, E.J. Copeland and N.J. Nunes, {\it Phys. Rev.} {\bf D61}, 127301 (1999).

\bibitem{Maju} A.S. Majumdar, {\it Phys. Rev.} {\bf D64}, 083503 (2001).

\bibitem{MTW} C.W. Misner, K.S. Thorne and J.A. Wheeler, {\it Gravitation}, W.H. Freeman and Company, 
              New York (1973).

\bibitem{alpha} J.K. Webb et al., {\it Phys. Rev. Lett.} {\bf 87}, 091301 (2001).

\bibitem{Jordan_1} P. Jordan, {Nature} {\bf 164}, 637 (1949); {\it Schwerkraft und Weltall}, Friedr. Vieweg \&
                   Sohn, Braunschweig (1952).

\bibitem{BD} C. Brans and R.H. Dicke, {\it Phys. Rev.} {\bf 124}, 925 (1961).

\bibitem{Wesson} P.S. Wesson, {\it Space, time, matter: Modern
              Kaluza-Klein theory}, World Scientific, 
              Singapore (1999).

\bibitem{HE} S.W. Hawking and G.F.R. Ellis, {\it The large scale
              structure of space-time}, Cambridge University Press, 
              Cambridge (1973).

\bibitem{NOOT} S. Nojiri, O. Obregon, S.D. Odintsov and V.I. Tkach
{\it Phys. Rev.} {\bf D64},043505 (2001).

\bibitem{Wett} C. Wetterich, {\it Nucl. Phys.} {\bf B302}, 668 (1988). 

\bibitem{Casimir} T. Appelquist and A. Chodos, {\it Phys. Rev.} {\bf D28}, 772 1983).

\bibitem{AW} J. Ambj\o rn and S. Wolfram, {\it Ann. Phys. (NY)} {\bf 147}, 33 (1983).

\bibitem{ABRS} A. Albrecht, C.P. Burgess, F. Ravndal and C. Skordis, {\it Phys. Rev.} {\bf D65}, 123506, 
               123507 (2002).

\bibitem{ENOO} E. Elizalde, S. Nojiri, S.D. Odintsov and S. Ogushi,
{\it hep-th/0209242}.

\bibitem{AS} A. Albrecht and C. Skordis, {\it Phys. Rev. Lett.} {\bf
84}, 2076 (2000).

\end{thebibliography}
\end{document}